\documentclass[manuscript]{an-fermipreprint}
\usepackage{graphicx}
\usepackage{times}
\usepackage{deluxetable}
\usepackage{threeparttable}
\newcommand{\angstrom}{\textup{\AA}}
\usepackage{natbib}
\bibpunct{(}{)}{;}{a}{}{,}

\begin{document}

% The following seven commands are intended for editorial usage and should be ignored by
% the author(s).
%\Pagespan{789}{}% Document's page range. 
% If second parameter is left empty, the last page is computed automatically.
%\Yearpublication{2017}%
%\Yearsubmission{2017}%
%\Month{05}%   
%\Volume{999}%  
%\Issue{88}% 
% \DOI{This.is/not.aDOI}% 

\title{Discovery of a New Quasar: SDSS J022155.26-064916.6}
%Example 
%for footnote, note the usage of the \texttt{fnmsep}
%command as separator between institute number and footnote mark} 
\author{Jacob M. Robertson\inst{1}\fnmsep\thanks{Corresponding author:
  \email{jrobertson15@my.apsu.edu}\newline}, Douglas L. Tucker\inst{2}, J. Allyn Smith\inst{1}, William Wester\inst{2}, Huan
  Lin\inst{2}, Jack H. Mueller\inst{2,3}, Deborah J. Gulledge\inst{1}%, Mees B. Fix\inst{3}
}
\titlerunning{Discovery of a New Quasar: SDSS J022155.26-064916.6}
\authorrunning{Jacob M. Robertson et~al.}
\institute{
Department of Physics \& Astronomy, Austin Peay State University, Clarksville, TN 37044
\and
Fermi National Accelerator Laboratory, Batavia, IL 60510-0500, USA
\and Illinois Mathematics \& Science Academy, Aurora, IL 60506-1000, USA
}

\received{2017 May 15}
\accepted{2017 May 15}
\publonline{2017 August 01 (expected)}


\abstract{%
We report the discovery of a new quasar: SDSS J022155.26-064916.6. This object was discovered while reducing spectra of a sample of stars being considered as spectrophotometric standards for the Dark Energy Survey. The flux and wavelength calibrated spectrum is presented with four spectral lines identified. From these lines, the redshift is determined to be $z \approx 0.806$. In addition, the rest-frame $u$-, $g$-, and $r$-band luminosity, determined using a k-correction obtained with synthetic photometry of a proxy QSO, are reported as $7.496 \times 10^{13}L_{\sun}$, $2.049 \times 10^{13}L_{\sun}$, and $1.896 \times 10^{13}L_{\sun}$, respectively.
} 

\maketitle

\section{Introduction}
As a probe into the distant and high-redshift universe, quasi-stellar objects (QSOs or quasars) are used in fundamental investigations into astrophysics and cosmology. These objects have refined our knowledge of the dynamics of supermassive blackholes \citep{Ghisellini} and the intergalactic medium \citep{Noterdaeme}. Furthermore, in cosmology, Quasars have been used to investigate the large-scale structure of the universe \citep{Springel,Croom} and the era of reionization \citep{Becker}. Through surveys such as the Sloan Digital Sky Survey (SDSS) \citep{York00} and the Two-Degree Fields Survey (2dF) \citep{Boyle}, thousands of these objects have been identified and studied. \par
We report the chance discovery of a new quasar, SDSS J022155.26-064916.6. The object (identification information shown in Table \ref{identify}) was imaged by SDSS and classified as a star with a cosmic ray hit, and it was not selected as a spectroscopic target for SDSS-III. Wide-field Infrared Survey Explorer (WISE) \citep{Wright10} data place the object in the QSO/Seyfert range, however, no spectroscopic confirmation of the object as a quasar previously existed. To support photometric calibrations for current and future sky surveys, such as the Dark Energy Survey (DES) \citep{Sanchez,Diehl14} and the Large Synoptic Survey Telescope (LSST) \citep{Tyson2008}, this object was selected as a potential DA (pure-hydrogen atmosphere) white dwarf star. While reducing data from this sample to verify target selection criteria, SDSS J022155.26-064916.6 was identified by its spectrum as a quasar.
The targeting, observations, data reductions (Section 2), and properties (Section 3) of this new quasar are described in the following sections. 

%\placetable{identify}

\begin{table}[h]
\centering
\tabletypesize{\footnotesize}
\rotate
\caption{Identification information for SDSS J022155.26-064916.6 
\label{identify}}
\tablewidth{0pt}
\begin{tabular}{cc}
\hline
ID/Coordinate & Value\\  
\hline
SDSS ID & J022155.26-064916.6 \\
SDSS photo ObjID & 1237679438814707830 \\
RA (J2000) & 35.4803 deg. \\
Dec (J2000) & -6.8213 deg. \\
\textit{l} & 173.664980849 \\
\textit{b} & -60.442953647 \\
\hline
GALEXASC & J022155.2-064916 \\
2MASS & J02215525-0649167 \\
WISE & J022155.26-064916.7 \\
\hline
\end{tabular}
\end{table}

\section{Target Selection, Observations \& Reductions}
Currently, a program is underway to construct a sample of DA white dwarfs in the DES footprint for use as spectrophotometric standards. This program supports DES calibrations using non-DES resources, such as photometric and spectroscopic data. Target selection for these standards was motivated by the SuperCOSMOS white dwarf survey \citep{RH11}, SDSS DR4 and DR7 white dwarf catalogs \citep{Eisenstein06,Kleinman13} respectively), and SDSS DR10 colors \citep{Ahn14}. Also contained in this sample was the quasar SDSS J022218.03-062511.1 previously identified during this program \citep{Fix}.

This object, a target for the candidate white dwarf follow-up program, was observed on 2013 December 9 using the Dual-Imaging Spectrograph (DIS) on the 3.5m ARC telescope at Apache Point Observatory (APO), New Mexico. Three 600 second exposures were collected with the DIS set to the standard low-blue/low-red configuration and using a 2.0" slit width. The DIS red channel contained the R300 grating which covers 4620 $\angstrom$ at 2.31 $\angstrom$/pix resolution. The DIS blue channel contained the B400 grating which covers 3660 $\angstrom$ at 1.83 $\angstrom$/pix resolution. Combined, the gratings cover 3600 to 9000 $\angstrom$. The data were processed using IRAF~\footnote{IRAF is distributed by the National Optical Astronomy Observatory, which is
operated by the Association of Universities for Research in Astronomy (AURA) under cooperative agreement
with the National Science Foundation.} spectroscopic packages. 

%\placetable{mags}
\begin{table}[h]
\centering
%\rotate
\caption{Photometric data for SDSS J022155.26-064916.6.
\label{mags}}
\tablewidth{0pt}
\begin{tabular}{cccc}
\hline
Filter & value & uncertainty & unit\\
\hline
GALEX-NUV & 18.32482 & 0.01344674 & Vega-mag \\
SDSS-u & 18.08 & 0.02 & AB-mag \\
SDSS-g & 17.80 & 0.01 & AB-mag \\
SDSS-r & 17.69 & 0.01 & AB-mag \\
SDSS-i & 17.67 & 0.01 & AB-mag \\
SDSS-z & 17.54 & 0.01 & AB-mag \\
2MASS-J & 17.284 & 0.108 & AB-mag\textsuperscript{*} \\
2MASS-J & 16.394 & 0.108 & Vega-mag \\
2MASS-H & 17.33 & 0.147 & AB-mag\textsuperscript{*} \\
2MASS-H & 15.960 & 0.147 & Vega-mag \\
2MASS-K & 17.26 & 0.169 & AB-mag\textsuperscript{*} \\
2MASS-K & 15.360 & 0.169 & Vega-mag \\
WISE-W1 & 16.369 & 0.025 & AB-mag\textsuperscript{*} \\
WISE-W1 & 13.686 & 0.025 & Vega-mag \\
WISE-W2 & 15.855 & 0.024 & AB-mag\textsuperscript{*} \\
WISE-W2 & 12.536 & 0.024 & Vega-mag \\
WISE-W3 & 15.171 & 0.044 & AB-mag\textsuperscript{*} \\
WISE-W3 & 9.929 & 0.044 & Vega-mag \\
WISE-W4 & 14.107 & 0.120 & AB-mag\textsuperscript{*} \\
WISE-W4 & 7.503 & 0.120 & Vega-mag \\

\hline

\end{tabular}
	\begin{tablenotes}
        \item{*}\footnotesize{The 2MASS and WISE magnitudes were converted to AB magnitudes by following 
        \cite{Odenwald03} and the description given on the WISE IPAC website
        (wise2.ipac.caltech.edu/docs/release/prelim/expsup/sec4\_3g.html)}
    \end{tablenotes}

\end{table}

\section{Results and Discussion}
\begin{figure}[h]
\includegraphics[scale=0.42]{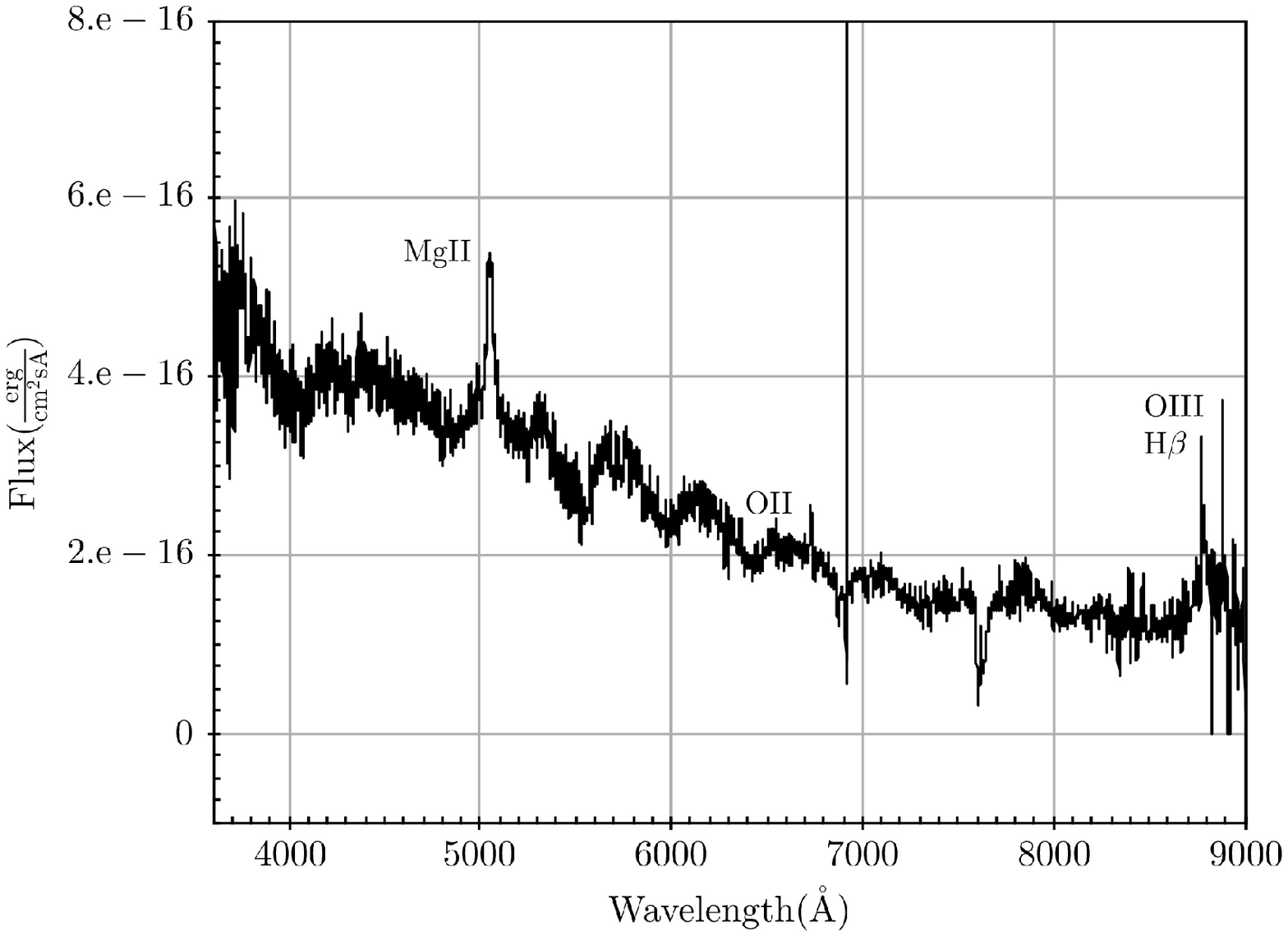}
\caption{A flux and wavelength calibrated spectrum of SDSS J022155.26-064916.6 in the redshift frame ($z=0.806$)
with labeled emission lines.  The line near 6900 $\angstrom$ is a processing artifact. \label{spectrum}}
\includegraphics[scale=0.42]{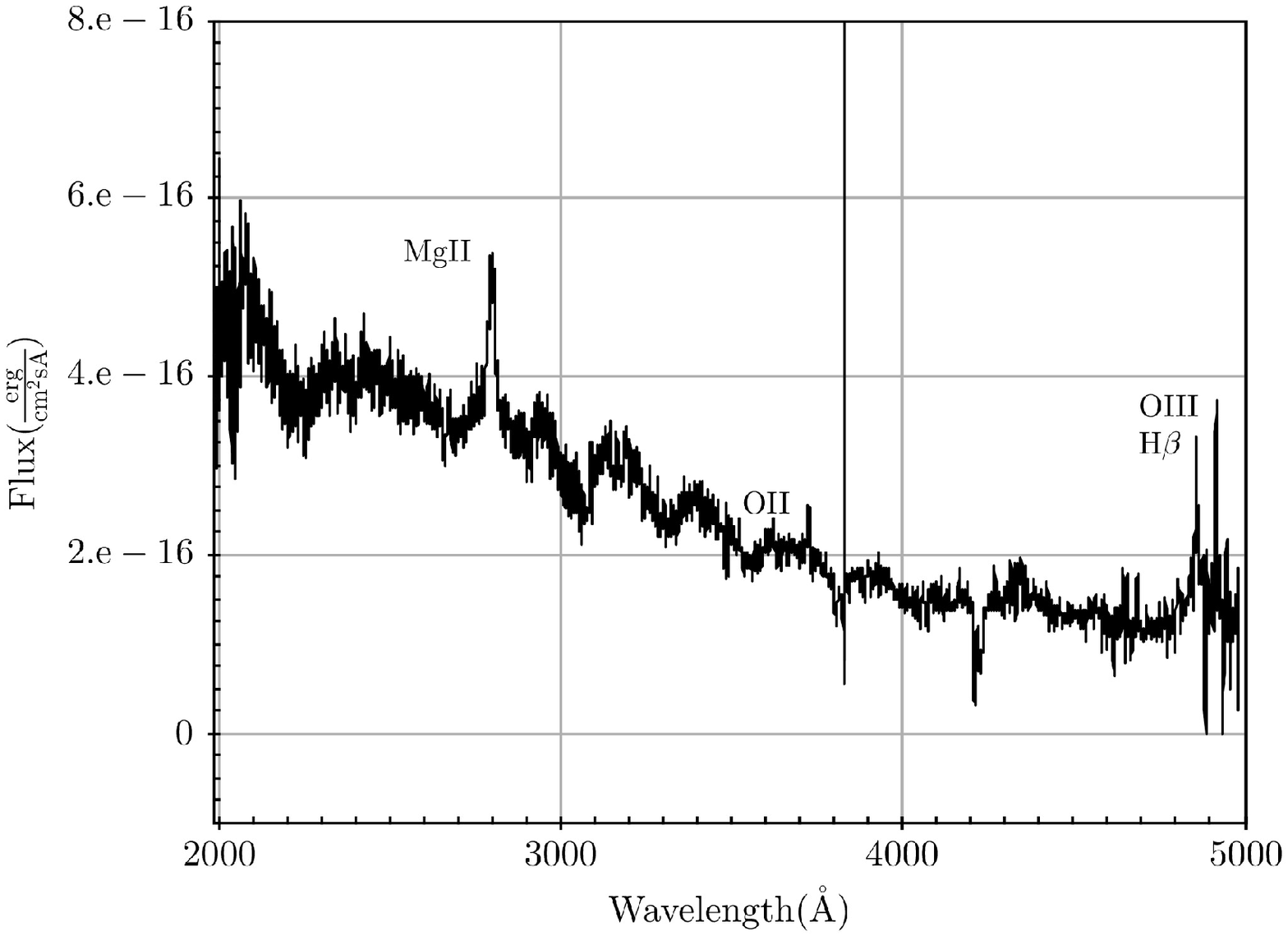}
\caption{A flux and wavelength calibrated spectrum of SDSS J022155.26-064916.6 in the rest  frame ($z=0.0$) 
with labeled emission lines.  The line near 3800 $\angstrom$ is a processing artifact. \label{spectra}}
\end{figure}

\begin{figure}[h]
\includegraphics[scale=0.42]{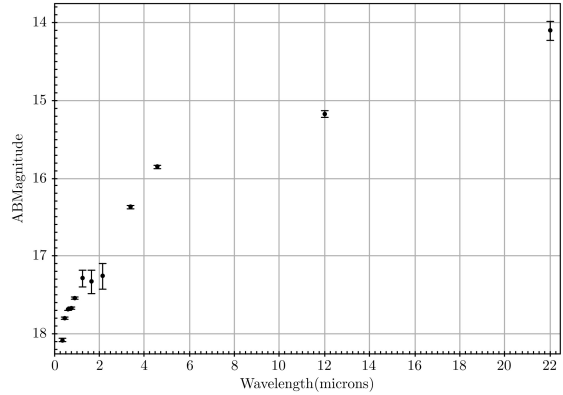}
\caption{The SED for SDSS J022155.26-064916.6 based on SDSS, 2MASS, and WISE photometric data (Table \ref{mags}). AB magnitudes are used; the 2MASS and WISE values were converted to AB values. \label{QSO-mags}}
\end{figure}

\begin{figure}[h]
\includegraphics[scale=0.45]{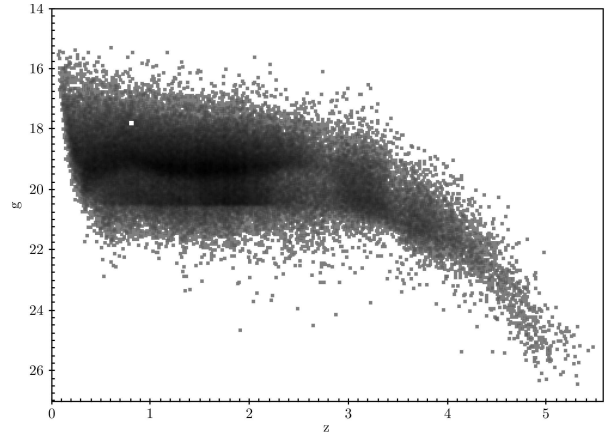}
\caption{The SDSS DR7 quasar catalog redshifts plotted against their $g$-magnitude. SDSS
J022155.26-064916.6 is shown as a yellow dot in relation to the other 100,000 objects in the
catalog. \label{redshift-g}}
\end{figure}

\begin{figure}[h]
\includegraphics[scale=0.45]{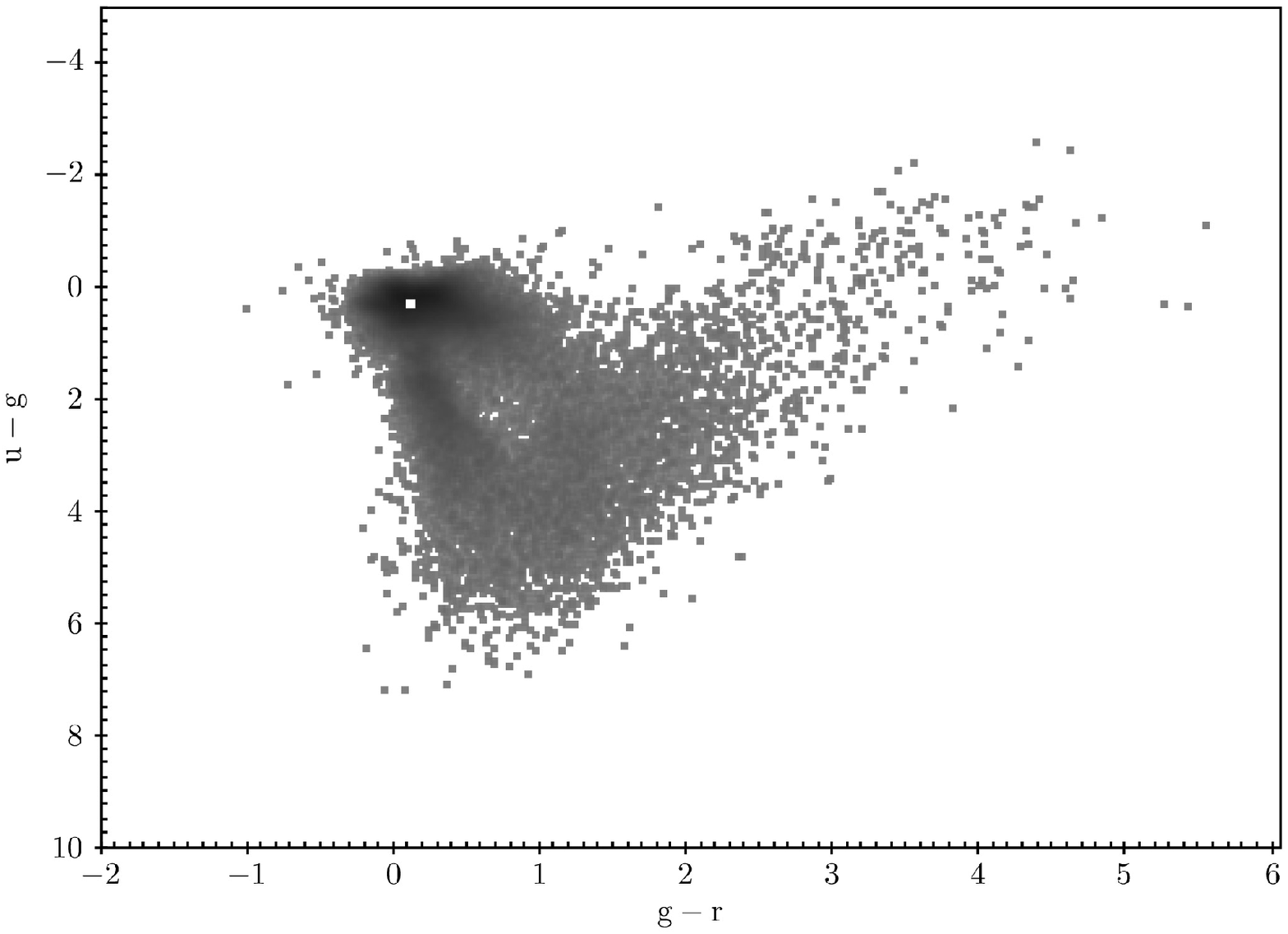}
\caption{The SDSS DR7 quasars shown in $(u-g)$ vs. $(g-r)$ space. SDSS J022155.26-064916.6is shown as a yellow dot in relation to the catalog data. \label{ug-gr}}
\end{figure}

\begin{figure}[h]
\includegraphics[scale=0.45]{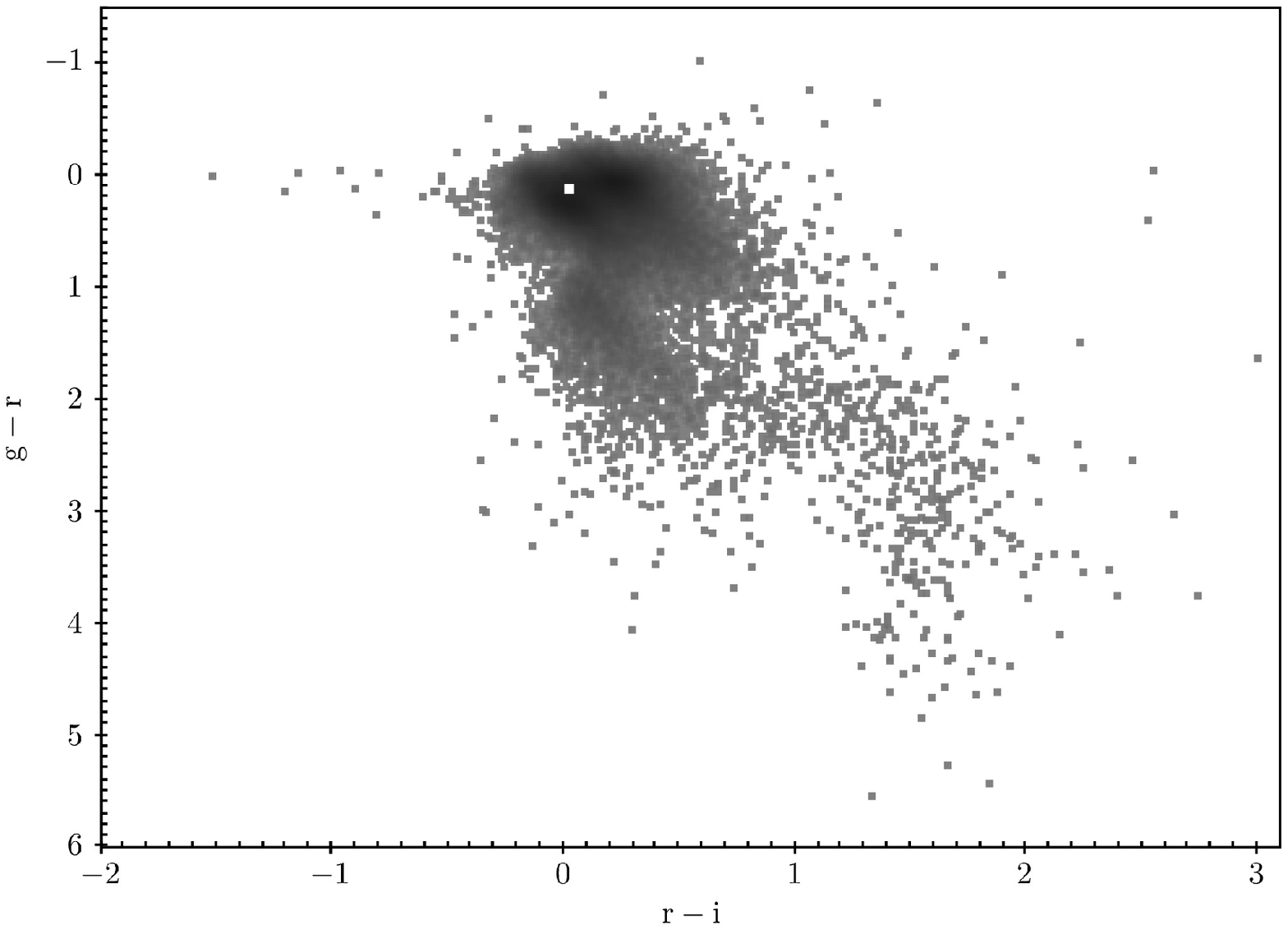}
\caption{The SDSS DR7 quasars shown in $(g-r)$ vs. $(r-i)$ space. SDSS J022155.26-064916.6 is shown as a yellow dot in relation to the catalog data. \label{gr-ri}}
\end{figure}

Investigation of the spectrum of SDSS J022155.26-064916.6, shown in Figures \ref{spectrum} and \ref{spectra}, consisted of identifying the emission lines with the \texttt{emsao} task in the IRAF external package \texttt{rvsao} \citep{Kurtz}. By interpreting the emission line around 5000 $\angstrom$ as MgII, the task returns $z=0.806$. Photometry from GALEX, SDSS, 2MASS, and WISE is shown in Table \ref{mags}. Following \cite{Odenwald03}, the 2MASS and WISE values were converted from Vega to AB values, keeping errors the same, for plotting in Figure \ref{QSO-mags}. Figure \ref{QSO-mags} shows the spectral energy distribution (SED) for SDSS J022155.26-064916.6 from SDSS, 2MASS, and WISE data. To compare this quasar to the quasars in the SDSS DR7 catalog, the object is plotted in Figure \ref{redshift-g} with the 100,000+ DR7 quasars in redshift vs. $g$-magnitude. Furthermore, Figures \ref{ug-gr} and \ref{gr-ri} compare $g$-$r$ vs. $u$-$g$ and $r$-$i$ vs. $g$-$r$, respectively. From these figures, we see that SDSS J022155.26-064916.6 is a fairly typical, moderately bright quasar.

The luminosity distance of SDSS J022155.26-064916.6 ($z\approx 0.806$) is found to be $D_{L}=5129.8$ Mpc ($H_{0}=69.9 \pm 0.7$ kms$^{-1}$; $\Omega_{M}=0.268\pm0.008$; $\Omega_{Vac}=0.714$; assuming a flat universe.) \citep{Wright2006,Bennett2014}.\footnote{http://www.astro.ucla.edu/$\sim$wright/CosmoCalc.html} Since complete coverage of the rest-frame and the observed-frame spectra does not exist in any SDSS filter, the k-corrections are determined for a proxy QSO (PG1100+772) from \cite{Shang}. The spectrum of PG1100+772, which covers the $u$-, $g$-, and $r$-bands, best matched the shape of SDSS J022155.26-064916.6, as shown in Figure \ref{PGspectrum}. Table \ref{abmagvalues} shows the k-correction \citep{Oke1968}, obtained using synthetic photometry of the rest and observed frames, for the SDSS $u$-, $g$-, and $r$-bands. The absolute magnitude for each filter band, using the respective interstellar extinction values from SDSS DR12 CAS\footnote{http://skyserver.sdss.org/dr12}, is also shown in Table \ref{abmagvalues}. 

\section{Conclusions}
A quasar was discovered serendipitously while reducing white dwarf spectra to support calibrations for the Dark Energy Survey. While WISE data place the object in QSO/Seyfert range, no spectroscopic confirmation of the object as a quasar previously existed.

%\placefigure{spectra}

%\placefigure{QSO-mags}

%\placefigure{abmagvalues}
\begin{table}
\centering
\tabletypesize{\footnotesize}
\rotate
\caption{The K-correction, interstellar extinction, and absolute magnitude values for SDSS J022155.26-064916.6 in $u$, $g$, and $r$. 
\label{abmagvalues}}
\tablewidth{0pt}
\begin{tabular}{cccc}
\hline
  value & $u$ & $g$ & $r$ \\  
\hline
k-correction & -0.3192 & -0.4306 & -0.1949  \\
extinction & 0.1384 & 0.1019 & 0.0739 \\
abs. mag & -28.137  & -28.159  & -28.514  \\
$L_{\sun}$ & $7.496 \cdot 10^{13}$ & $2.049 \cdot 10^{13}$  & $1.896 \cdot 10^{13}$\\ 

\hline
\end{tabular}
\end{table}

%\placefigure{redshift-g}

%\placefigure{ug-gr}

%\placefigure{gr-ri}

%\placefigure{PGspectrum}
\begin{figure}[h]
\label{PGspectrum}
\includegraphics[scale=0.35]{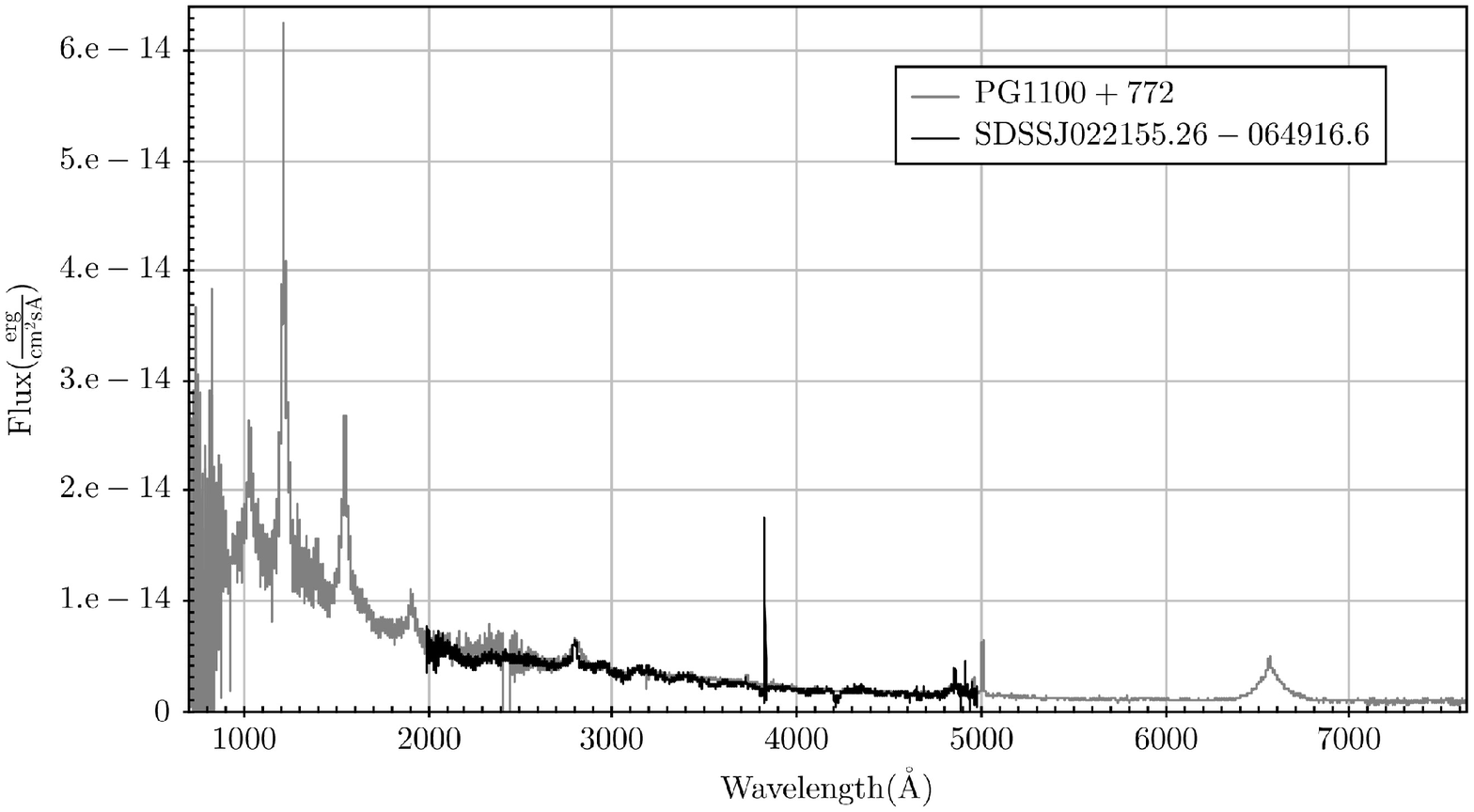}
\caption{SDSS J022155.26-064916.6 and PG1100+772 plotted in the rest frame ($z=0.0$). The flux for SDSS J022155.26-064916.6 was multiplied by 12. \label{PGspectrum}}
\end{figure}

\section{Acknowledgements}
Partial support for JMR was provided by the DES operations group for calibration support.  Additional support came from the Fermilab
Center for Particle Astrophysics. Based on observations obtained with the Apache Point Observatory
3.5-meter telescope, which is owned and operated by the Astrophysical Research Consortium.

Fermilab is operated by Fermi Research Alliance, LLC under Contract No. De-AC02-07CH11359 with the United States Department of Energy.

Funding for SDSS-III has been provided by the Alfred P. Sloan Foundation, the Participating
Institutions, the National Science Foundation, and the U.S. Department of Energy Office of Science.
The SDSS-III web site is http://www.sdss3.org/.

SDSS-III is managed by the Astrophysical Research Consortium for the Participating Institutions of
the SDSS-III Collaboration including the University of Arizona, the Brazilian Participation Group,
Brookhaven National Laboratory, Carnegie Mellon University, University of Florida, the French
Participation Group, the German Participation Group, Harvard University, the Instituto de
Astrofisica de Canarias, the Michigan State/Notre Dame/JINA Participation Group, Johns Hopkins
University, Lawrence Berkeley National Laboratory, Max Planck Institute for Astrophysics, Max
Planck Institute for Extraterrestrial Physics, New Mexico State University, New York University,
Ohio State University, Pennsylvania State University, University of Portsmouth, Princeton
University, the Spanish Participation Group, University of Tokyo, University of Utah, Vanderbilt
University, University of Virginia, University of Washington, and Yale University.

This research has made use of the NASA/IPAC Extragalactic Database (NED), which is operated by the
Jet Propulsion Laboratory, California Institute of Technology, under contract with the National
Aeronautics and Space Administration. This research has made use of the SIMBAD database, operated
at CDS, Strasbourg, France. This research made use of data from the GALEX mission.  GALEX is a NASA
small explorer, launched in 2003 April. It is operated for NASA by Caltech under NASA contract
NAS5-98034.  This publication makes use of data products from the Wide-field Infrared Survey
Explorer, which is a joint project of the University of California, Los Angeles, and the Jet
Propulsion Laboratory California Institute of Technology, funded by the National Aeronautics and
Space Administration.

The TOPCAT software package\footnote{http://www.starlink.ac.uk/topcat/} was the plotting tool for this work.

\newpage

\end{document}